# Redefining climate index with sea surface temperature of key area can better express atmospheric physical activity

Jishi Zhang*, Jingrong Lu, Haidong Huang, Xiaoning Liu, Yafeng Zhao, Zhiyang Wang, Jinlei Mo

（*School of Environmental and Municipal Engineering, Lanzhou Jiaotong University,Lanzhou 730070, China*）

**Abstract** Since the 1920s, when the three oscillations ENSO, NAO and NPO were found and defined, the theory of ocean-atmosphere coupling is particularly important in explaining the phenomena of atmospheric physics, then defined more than ten climate indexes, such as SOI, AO, AAO and Nino monitoring area and so on, the indexes usually adopt standardized sea level pressure (SLP) difference or standardized sea surface temperature anomaly (SSTA) difference definitions. The survey is carried out with seventy stations runoff of major rivers in the world shows that the correlation coefficient between these indexes and runoff is very low, most of which is between 0.1-0.2, the impact of the index on the hydrologic climate cannot be truly reflected, which has led to doubts about the clear physical mechanism and even questions of teleological authenticity. This is due to the fact that standardization is a nonlinear process, which changes the structure of the original data, causing the correlation coefficient to be low. Using sea surface temperature difference (SSTD) to define these climate indexes, the correlation coefficient is significantly improved, the Hankou station of Yangtze river, Darband station of Ind, Hope station of Fraser, Bangui station of Oubangui generally exceeds 0.8 strong degree, which prove that these relevant elements have a high resonance effect, there must be a common driving physical quantity. We find that the sun-earth distance has a very strong resonance effect, it is the earth's rotational orbit that determines their common activity intensity and cycle characteristics. The sun-earth distance can accurately calculate the tens and hundreds years of SST and SSTD, if using the SST or SSTD as the climate indicators, which express physical process of various ocean-atmosphere coupling, we can thoroughly solve long-term ultra-long-term forecast of international difficulty problems.

**Keywords** Sea surface temperature difference; correlation coefficient; Climate index

## 1. Introduction

Ocean-atmosphere coupling is an important atmospheric physical process affecting the global hydrologic climate, which has attracted widespread attention. Walker (1924; 1932) saw the seesaw phenomenon of atmospheric activities center and system proposed the southern oscillation (SO), the north Atlantic oscillation (NAO), and the north Pacific oscillation (NPO) three oscillations concept, it essentially reflects atmospheric circulation phenomenon caused by the thermal status differences, which profound impact on the development of climatology. Bjerknes (1966; 1969) found SO and El Nino reflect the two expression forms of the tropical oceanic and atmospheric movement, the physical mechanism is the same, and introduce the sea surface temperature (SST) in the equatorial Pacific climate researches for the first time. ENSO, as global scale ocean-atmosphere coupling phenomenon, continues to be focus of researchers. Walker (1924; 1932) defined the SOI with sea level pressure (SLP), temperature, precipitation and various meteorological elements, early was used to reflect the inverse relationship of air pressure between the Pacific and the Indian (Wright 1977; George et al. 1989). Although this definition in observation fact and the physical mechanism interpretation is very perfect, in practice, the correlation coefficient between SOI and various climate hydrological sequences is very low, people mistakenly think that is teleconnection (Kiladis and Diaz 1989; Rasmussen and Wallace 1983; Trenberth and Hoar 1996; McCABE and Dettinger 1999). Since then, a variety of new definitions of SOI have been proposed by studying ENSO (Troup 1965; Wright 1975; Trenberth 1976; McBride and Nicholls 1983), try to better describe ENSO. The U.S. climate analysis center (CAC) standardized difference of air pressure anomaly between Tahiti and Darwin to define SOI. Wolter (1998, for further reference) defined the multivariate ENSO index (MEI) using six variables on the Pacific found that MEI is a good indicator of measuring ENSO events. Observational facts and actual climate diagnosis indicate that sea surface temperature anomaly (SSTA) signal is the most obvious in eastern equatorial the Pacific, so the tropical Pacific is divided from four SST key areas, these are Nino1+2, Nino3, Nino4 and Nino3.4 as indicators to reflect EI Nino and La Nina event, and the standard anomaly difference index is given, but the correlation with climatic variation is still very low. So researchers suspect that may exist other SST key area influencing the global climate, there are general researches on the ocean with the method of empirical orthogonal function (EOF), and defines the arctic oscillation (AO), the Antarctic oscillation (AAO), the Indian Ocean dipole (IOD), India-Pacific(IP) joint oscillation, equatorial Atlantic-Pacific(EAP) joint oscillation and so on all kinds of ocean-atmosphere coupling theories. Saji (1999)studied the phenomenon of SST positive and negative anomaly in western and eastern Indian Ocean, and named the phenomenon as the IOD, the equatorial IOD has an important effect on atmospheric circulation, especially in the low latitude of the eastern hemisphere (Webster et al 1999).The IOD is connected to ENSO through Indonesian through flow, walker circulation and tropical atmospheric bridges (Stephen et al 1999; Wyrtki 1961). Latif (1995) proposed the Atlantic and Pacific Walker circulation anomaly mechanisms. Wang (2006) defines an EAP SST zonal gradient index with interactions between the two ocean basins. Ju Jianhua (2004) proposed the concept of the IPSSTA synthetic modal and defined a synthetic modal index. The IP ocean-atmosphere system has a big influence on global weather and climate.

Thompson and Wallace (1998) found that high latitudes and low latitudes SLP between the two air ring active belt is a global scale seesaw phenomenon in the northern hemisphere, this atmospheric circulation control the basic situation of atmospheric movement, is an important factor influencing the weather and climate characteristics of the northern hemisphere, on basis of the feature proposed the AO. Montgomery (1940) based on analysis of a small number of site observations, suggests that there may be also a new oscillation in the high and mid-latitudes of the southern hemisphere. Kidson (1975) conducted the analysis of the main components of SLP data from 1951 to 1960 in the southern hemisphere. Roger (1982) carried on the EOF analysis, the spatial distribution characteristic of main feature vector is around 50°S bounded, the south is the opposite the north, the Antarctic continent is extreme center. So Gong (1999) defined the average SLP difference between 40°S and 65°S as the AAO, which well represents the characteristics of the AAO. At the same time, people also name the AO and AAO as annular mode, and praised the two oscillations, considered the SO, AO and AAO control of the global atmospheric circulation situation.

Madden and Julian (1971) defined the interaction between the Indian and Pacific zonal wind and air pressure as the Madden-Julian oscillation (MJO), the tropical wind field show the low-frequency oscillation of the 30~60d cycle of the atmospheric pressure structure, Matthews (2004) pointed out that the ocean-atmosphere interaction was the main factor of MJO, Zhang (2005) believed that MJO is a large-scale perturbation of the atmospheric deep convection. Xue Y. (2016) found that every WWB event outbreak and gradually eastward expansion was associated with MJO. Lin (2009) pointed out that the convective anomaly associated with MJO reaches the equatorial Indian and the western Pacific region after 5~15d, the amplitude of NAO will increase, and certain phase of MJO ahead NAO events.

Atmospheric oscillation indexes are basically defined by difference of SLP or SSTA standardization, essentially reflect the atmospheric movement phenomenon caused by the earth thermal status differences. SST is one of the basic characteristics of the thermal difference of the earth, is affected by the sea internal and external environment, its changes produce further impact on the atmosphere, so the SST plays a vital role in the atmospheric circulation. Over the years, people continue to in-depth study, build all kinds of theory system, at the same time actively seeking new oscillation index to improve correlation coefficient of global hydrology and climate physical parameters. In order to significantly increase the correlation coefficient of the oscillation index and hydrological climate data, and make qualitative analysis in the global climate forecast. In this paper, on the basis of predecessors' research attempts to directly use the key area SST or SSTD definition oscillation index, and use the existing early global runoff data to perform simple validation. Main consideration related statistics commonly is very simple and reliable calculation method, we construct climate indicators mainly adopt the reanalysis data, which has nothing to do with global runoff data more than 30 years ago, to avoid unnecessary argument.

**2 Data and methodology**

2.1 Datasets

The southern oscillation index(SOI), MEI, Nino1+2, Nino3, Nino4, Nino3.4 and the NPO are from https://www.esrl.noaa.gov/psd/data/climateindices/list;The AO is from https://www.cpc.ncep.noaa.gov;The NAO is from https://www.esrl.noaa.gov/psd/gcoswgsp/timeseries;The AAO is from https://www.esrl.noaa.gov/psd/data/20thC_Rean/timeseries/monthly/AAO/; Global SST dates that extended reconstruction gridded at a 2°×2° resolution between 1840 and 2010 are derived from the national oceanic and atmospheric administration (NOAA) (http://www.noaa.gov/).

The global runoff data as verified with various climate indicators is from the State Hydrological Institute (http://www.hydrology.ru/) (Table 1).

Tab. 1 Information of hydrological site

| Continent | River | Station | Latitude (deg.) | Longitude (deg.) | Area (km$^2$) | Year | Continent | River | Station | Latitude (deg.) | Longitude (deg.) | Area (km$^2$) | Year |
|---|---|---|---|---|---|---|---|---|---|---|---|---|---|
| Asia | Yangtze River | Hankou | 30.58 | 114.28 | 1488036 | 1866-1986 | Africa | Blue Nile | Khartoum | 15.61 | 32.55 | 325000 | 1900-1937 |
| | Yangtze River | Datong | 30.76 | 117.61 | 1705383 | 1947-1986 | | Oubangui | Bangui | 4.37 | 18.58 | 500000 | 1936-1975 |
| | Yellow River | maqu | 33.96 | 102.09 | 86048 | 1959-2001 | | Sangha | Ouesso | 1.61 | 16.05 | 158350 | 1947-1983 |
| | Yellow River | Longmen | 35.4 | 110.35 | 497552 | 1934-1979 | North America | Fraser | Hope | 49.38 | -121.45 | 217000 | 1913-1984 |
| | Songhuajiang | Haerbin | 45.77 | 126.58 | 391000 | 1898-1983 | | Columbia | The Dalles | 45.6 | -121.17 | 613830 | 1879-1984 |
| | Xijiang | Wuzhou | 23.48 | 111.3 | 329705 | 1941-1984 | | Ohio | Metropolis | 37.13 | -88.73 | 525770 | 1928-1984 |
| | Ind | Darband | 34.4 | 72.83 | 166000 | 1937-1970 | | Missouri | Hermanmo | 38.7 | -91.43 | 1357677 | 1897-1984 |
| | Godavary | Davlaishwaram | 16.55 | 81.47 | 299300 | 1902-1974 | | Mississippi | Vicksburg | 32.3 | -90.9 | 2964254 | 1928-1982 |
| | Krishna | Vijayawada | 16.52 | 80.62 | 251355 | 1901-1979 | | Mississippi | Alton | 38.88 | -90.17 | 444185 | 1928-1984 |
| | Ob | Belogore | 61.07 | 68.6 | 2160000 | 1936-1989 | South America | Amazonas | Obidos | -1.9 | -55.5 | 4640000 | 1928-1983 |
| | Enisey | P.Tunguska | 61.6 | 90.08 | 1760000 | 1936-1989 | | Parana | Guaira | -24.06 | -54.26 | 802150 | 1920-1979 |
| Europe | Lena | Kyusyur | - | - | 2430000 | 1936-1994 | | S.Francisco | Juazeiro | -9.41 | -40.5 | 510800 | 1929-1979 |
| | Volga | Volgograd | 48.7 | 44.52 | 1350000 | 1879-1935 | | Magdalena | Calamar | 10.27 | -74.92 | 257438 | 1971-1990 |
| | Danube | D-T Severin | 44.7 | 22.42 | 576232 | 1866-1988 | Australia & Oc. | Murray | Albury | -36.1 | 146.9 | 17220 | 1877-1950 |
| Africa | Nile | Dongola | 19.18 | 30.48 | - | 1912-1984 | | Murray | Swan Hill | -35.33 | 143.57 | - | 1909-1966 |
| | White Nile | Malakal | 9.58 | 31.62 | 1080000 | 1912-1982 | | Waikato | Taupo'Control | 38.7 | 176.07 | 3290 | 1906-1967 |

Definition areas of various oscillation indexes are determined on the basis of predecessors' in-depth analysis calculation, we redefined average SSTD between the high temperature and the low temperature area with predecessors determined area (table 2).

Table 2 definition sea area of the oscillation index

| Index | Area | Longitude(deg.) | Latitude(deg.) | Year | Index | Area | Longitude(deg.) | Latitude(deg.) | Year |
|---|---|---|---|---|---|---|---|---|---|
| ENSOSSTD | Tahiti | 130°W~155°W | 4°N~10°N | 1870-2010 | NAOSSTD | Azores | 26°W~32°W | 36°N~40°N | 1854-2010 |
| | Darwin | 90°W~150°W | 4°S~4°N | | | Iceland | 18°W~26°W | 62°N~66°N | |
| WIPSSTD | Indian | 50°E~70°E | 10°S~10°N | 1854-2010 | NPOSSTD | Hawaii | 130°E~170°E | 24°N~34°N | 1854-2010 |
| | Pacific | 130°W~155°W | 4°N~10°N | | | Aleutian | 144°~170°E | 44°N~56°N | |
| APSSTD | Atlantic | 20°E~45°E | 0°~25°N | 1854-2010 | IODSSTD | W. Indian | 50°~70°E | 10°S~10°N; | 1854-2010 |
| | Pacific | 90°W~165°W | 4°S~4°N | | | E. Indian | 90°N~110°E | 10°S~0° | |
| AOSSTD | 50ºN Circle | 0º~360ºW | 50ºN | 1854-2010 | AAOSSTD | 40ºS Circle | 0º~360ºW | 40ºS | 1854-2010 |
| Nino1+2 | | 80˚W~90˚W | 0˚S~10˚S | 1854-2010 | Nino 4 | | 150˚W~160˚W | 5˚S~5˚N | 1854-2010 |
| Nino 3 | | 90˚W~150˚W | 5˚S~5˚N | 1854-2010 | Nino 3.4 | | 120˚E~170˚W | 5˚S~5˚N | 1854-2010 |

2.2 Methodology

Meteorology and hydrology are different stages in the process of the atmosphere heat and moisture physical cycle, although the mechanism that drives their physical movement is consistent, their response time have own sequence, particularly in the far distance it is hard to show up at the same time. Therefore, it is necessary to carry out delay and advance correlation analysis. The time series was ordered by month, and the correlation analysis was carried out month by month, the range of previous and delay time is 0-12 months. The formula for correlation analysis as follow:

$$r_{xy} = \frac{\sum_{i=1}^{n}(x_i - \bar{x})(y_i - \bar{y})}{\sqrt{\sum_{i=1}^{n}(x_i - \bar{x})^2} \sqrt{\sum_{i=1}^{n}(y_i - \bar{y})^2}} \quad i = 1,2,3,\ldots, n \pm m$$

Where, $n$ is the total month; and $m$ is the previous or delay months, $i = 1,2,3,\ldots, n \pm m$; $m$=-12,…,3,-2,-1,0,1,2,3,…,12. Minus refers to previous and plus refers to delay. In essence, delay and advance correlation is the same, because of the hydrology and climate is controlled by the revolution of the earth, the influence extent is cycle changed, advance correlation can test the stability of the driving factor circulation.

The 95% confidence test of the correlation coefficient, the lowest critical correlation coefficient $r_k$ is calculated as:

$$r_k = \sqrt{\frac{t_\alpha^2}{n - 2 + t_\alpha^2}}$$

$n$ is greater than 120 when $t_{0.05} = 1.980$, $r_k$= 0.179, the correlation coefficient greater than or equal to 0.179 through the test, otherwise no pass the test; n greater than 1200 $t_\alpha$ = 1.906, $r_k$ = 0.055, correlation coefficient greater than or equal to 0.055 through the test, otherwise no pass test.

**3 Various southern oscillation indexes**

Ocean-atmosphere coupling is a system with interactive coupling and resonance effect, the observation and numerous studies show that the global low and mid-latitudes ocean-atmosphere coupling variation is an organic whole, no matter whether from theinterannual and interdecadal exist coupling synchronization. When EI Nino event occurs, the SST changes often begin from the coast of Peru and Ecuador (Nino1+2 area), and then expand to the eastern and equatorial Pacific. Jean-Philippe (2005) believes that because of precipitation pattern of El Nino and La Nino events is asymmetrical, which make Nino3.4 or Nino1+2 index is non-linear relationship with precipitation, in the case of linear statistical forecast of regional precipitation, the effect is not good. McLean (2009) investigated the impact of the SO on global temperature is evident, especially in the equatorial region, which account tropical temperature anomalies for 81% of the variance.

3.1 SOI

The general use of SOI is defined by the mean SLP difference between Papeete and Darwin, its specific definition formula is,

$$SOI = 10 \times \frac{SLPdiff - avSLPdiff}{StdDev(SLPdiff)}$$

where SLPdiff is sea-level pressure difference between Papeete and Darwin for the month, avSPLdiff is the annual average value of the difference, and StdDev(SLPdiff) is standard deviation. Although SOI has good application effects in the macroscopic description of the ENSO definition and global climate response, many studies showed that correlation of the SOI and river runoff, local rainfall and temperature change is weak (Jin et al. 2005), most of the correlation coefficients are below 0.1, which lead to ENSO event is controversial in macroscopic qualitative research. Based on correlation analyses of SOI and global runoff, there is a remote correlation of SOI and global runoff, most hydrological stations correlation coefficient are less than 0.1, the highest correlation coefficient is only 0.32(Tab.3), which is in line with a number of related research conclusion. This is the disadvantage of standardized index, due to the Standardized processing is a nonlinear, changed the structure of the original data, so the correlation coefficient is very low, which is the improper mathematical expression or use of the index.

Tab.3 The correlation coefficient between the SOI, MEI, SSTD and global runoff

| Station | Max correlation coefficient(month) | | | Station | Max correlation coefficient(month) | | |
| --- | --- | --- | --- | --- | --- | --- | --- |
| | SOI | SSTD | MEI | | SOI | SSTD | MEI |
| Hankou | 0.06(-8) | -0.67(0) | -0.09(-9) | Khartoum | 0.15(2) | 0.56(-4) | -0.15(-8) |
| Datong | 0.10(-10) | -0.74(12) | -0.13(-5) | Bangui | -0.07(12) | -0.72(-10) | 0.07(5) |
| maqu | -0.09(-11) | -0.64(-11) | -0.07(7) | Ouesso | 0.05(-9) | -0.64(+2) | -0.08(-12) |
| Longmen | 0.07(+4) | -0.62(+1) | -0.09(-3) | Hope | 0.06(5) | -0.68(-1) | -0.07(8) |
| Haerbin | 0.04(-9) | -0.54(+1) | -0.06(-12) | The Dalles | 0.10(5) | -0.62(-2) | -0.13(5) |
| Wuzhou | 0.05(-7) | -0.69(12) | -0.07(-4) | Metropolis | 0.11(1) | 0.50(2) | -0.09(2) |
| Darband | 0.07(0) | -0.71(-) | -0.07(-4) | Hermanmo | -0.10(-9) | -0.45(-3) | -0.11(-12) |
| Davlaishwaram | 0.06(-7) | -0.61(-11) | -0.08(-4) | Vicksburg | 0.15(-10) | -0.52(-3) | -0.12(-9) |
| Vijayawada | 0.06(2) | -0.62(-11) | -0.08(-3) | Alton | 0.11(-10) | -0.49(9) | 0.11(12) |
| Belogore | -0.05(7) | -0.66(-1) | 0.05(12) | Obidos | 0.06(+4) | 0.62(-9) | -0.11(5) |
| P.Tunguska | 0.05(-8) | -0.58(11) | 0.06(12) | Guaira | -0.08(-5) | -0.60(-6) | 0.12(-3) |
| Kyusyur | 0.05(1) | -0.53(11) | -0.05(-2) | Juazeiro | -0.10(-8) | -0.63(-5) | 0.09(-5) |
| Volgograd | -0.07(9) | -0.58(10) | -0.05(-12) | Calamar | 0.32(1) | 0.59(8) | -0.32(2) |
| D-T Severin | -0.11(1) | -0.50(9) | 0.13(6) | Albury | 0.19(3) | -0.47(-10) | -0.17(-1) |
| Dongola | 0.09(-5) | -0.64(+2) | -0.11(-2) | Swan Hill | 0.21(+4) | 0.52(+8) | -0.22(1) |
| Malakal | 0.13(+6) | 0.60(+8) | -0.14(3) | Taupo'Control | 0.12(1) | 0.29(9) | 0.11(-12) |

Note:"()"indicates that the month of advance or delay, positive is the delay and the negative is advance.

3.2 MEI

In 1993 the U.S. climate diagnosis center (CDC) put forward the ENSO index (MEI) composed of six major Pacific Ocean observation variable multiple, which is better able to respond to the heat factor of the sea-air system. The six variables include: SLP, surface wind of latitude and longitude, SST, sea surface air temperature and air total cloud cover, so the MEI is joint index of six rotating principal component observation area .The result shows that the correlation coefficient between MEI and global runoff is about 0.1, the maximum is 0.32, compared to the correlation between SOI and global runoff, the coefficient increased slightly, no substantial increase, still cannot meet the needs of the local hydrology and climate numerical analysis (table 3).

3.3. ENSOSSTD

As mentioned earlier, the ENSO phenomenon plays an important role in the study of hydrology and climate, it is necessary to solve the problem that the correlation between ENSO and local hydrologic climate is extremely low. The global ocean-atmosphere circulation is essentially reflection of the thermal condition differences, thus directly using the SSTD should be able to better reflect the correlation between the ocean-atmosphere movement and the local hydrology and climate. SST are is an important indicator used to describe the marine physical characteristics, early people found that the cold SST and hot reverse phase change in eastern and the western Pacific, which was defined the ENSO phenomenon. Due to early global large-scale joint research rarely, shortage of SST data and meteorological data, after the global SST data and climatic information assimilation are improved, greatly improve the accuracy and reliability, with reprocessing SST to redefine the ENSO index into the possible. We select average SSTD of Tahiti near cold waters and Darwin near warm waters to define the ENSOSSTD. Correlation analysis between ENSOSSTD and global runoff see Tab.3, the results show that correlation coefficient of all the stations almost over 0.1, most of the stations correlation coefficient range from 0.5 to 0.74, the maximal correlation coefficient is 0.74. Therefore, it can be seen that SSTD can better reflect the influence of the SOI on the global climate, and also prove the reliability and rationality of the reanalysis data from the other side.

3.4 Nino index

Due to a large number of studies have found that correlation coefficient between SOI and the local hydrology and climate site is low, but the physical mechanism expressed by the ENSO phenomenon is very clear, which is not matched with the extremely low correlation coefficient. People try to improve the correlation coefficient by modifying or redefining the ENSO index. Nino C SST, Nino3 and Nino3.4 SST have all been used for more accurate indicators to reflecting ENSO (Angell 1981; Wang and Gong 1999; Trenberth 1997). The correlation analysis of Nino monitoring area anomaly index and global runoff is generally low, and the maximum is only 0.35, is not good enough to characterize the effect of ENSO phenomenon on global climate (Tab.4). Using the average SST of the four regional as Nino1+2, Nino3, Nino4 and Nino3.4 SST index to verify with global runoff, the correlation between Nino SST index and global runoff has significantly improved, especially Nino1+2 performance is particularly prominent, the largest correlation coefficient can reach 0.84; Nino4 has the lowest correlation coefficient, it essentially reflects MJO.

Tab.4 The correlation coefficient between the Nino indexes and global runoff

| Station | Max correlation coefficient(month) | | | | | | | |
|---|---|---|---|---|---|---|---|---|
| | Nino1+2SSTA | Nino1+2SST | Nino3SSTA | Nino3SST | Nino4SSTA | Nino4SST | Nino3.4SSTA | Nino3.4SST |
| Hankou | -0.10(-5) | -0.79(+1) | -0.09(-9) | -0.68(+2) | -0.18(-1) | -0.31(+5) | -0.10(-11) | -0.47(+3) |
| Datong | 0.19(12) | -0.79(-2) | 0.13(9) | -0.69(-2) | -0.21(-2) | -0.35(-5) | 0.14(7) | -0.51(-4) |
| maqu | -0.07(7) | 0.68(-7) | 0.05(-12) | 0.57(-8) | -0.14(0) | -0.29(7) | -0.08(1) | 0.38(-9) |
| Longmen | -0.18(-3) | -0.59(+1) | -0.11(3) | 0.5(+8) | -0.15(0) | -0.30(+5) | -0.13(0) | 0.35(+8) |
| Haerbin | 0.09(1) | 0.65(-5) | -0.10(-11) | 0.56(-4) | -0.16(-12) | -0.22(+5) | -0.12(-12) | 0.36(-4) |
| Wuzhou | -0.12(-5) | 0.74(-8) | -0.07(-4) | 0.62(3) | 0.14(4) | -0.28(-6) | -0.09(-3) | 0.44(2) |
| Darband | -0.13(-5) | 0.82(-8) | -0.10(-3) | 0.73(-9) | -0.17(-2) | -0.33(-6) | -0.12(-2) | 0.53(-10) |
| Davlaishwaram | -0.11(-4) | 0.70(-7) | -0.09(-3) | 0.59(-8) | -0.15(-2) | -0.31(-5) | -0.10(-1) | 0.41(-9) |
| Vijayawada | -0.10(-3) | 0.72(-7) | -0.09(0) | 0.61(-8) | -0.16(-1) | -0.31(-5) | -0.10(0) | 0.43(-9) |
| Belogore | -0.08(-6) | 0.77(4) | -0.06(-12) | 0.64(3) | 0.13(4) | 0.26(12) | 0.09(5) | 0.45(2) |
| P Tunguska | 0.10(12) | 0.71(3) | 0.07(5) | 0.58(2) | 0.11(3) | 0.26(12) | 0.10(5) | 0.41(1) |
| Kyusyur | -0.09(-6) | 0.67(4) | -0.03(1) | 0.55(-9) | 0.09(4) | 0.23(-11) | -0.06(0) | 0.40(2) |
| Volgograd | 0.08(8) | 0.69(3) | -0.08(-11) | 0.60(2) | 0.17(3) | 0.26(0) | 0.08(5) | 0.44(1) |
| D-T Severin | 0.12(7) | 0.51(2) | 0.15(5) | 0.48(1) | 0.20(2) | 0.20(-1) | 0.17(4) | 0.36(0) |
| Dongola | -0.17(-3) | -0.68(0) | -0.11(3) | 0.60(+7) | -0.18(0) | -0.31(+5) | -0.12(2) | 0.43(+8) |
| Malakal | -0.19(10) | -0.73(-1) | -0.14(4) | -0.63(0) | -0.24(1) | 0.27(+9) | -0.17(2) | -0.42(+1) |
| Khartoum | -0.10(1) | 0.68(-6) | -0.14(-7) | 0.54(-7) | -0.17(-1) | -0.34(-5) | -0.15(-8) | -0.38(-2) |
| Bangui | -0.13(-2) | 0.84(7) | 0.07(7) | 0.75(6) | 0.17(6) | 0.27(4) | 0.12(7) | 0.54(5) |
| Ouesso | -0.15(10) | -0.68(-1) | -0.08(6) | 0.59(+6) | -0.13(-12) | -0.29(-9) | 0.07(-3) | 0.41(+7) |
| Hope | -0.12(6) | 0.78(-8) | -0.06(12) | 0.67(-10) | -0.16(10) | -0.30(5) | 0.10(-7) | 0.49(-10) |
| The Dalles | -0.12(5) | 0.73(-9) | -0.11(2) | 0.63(-10) | -0.20(8) | -0.35(4) | -0.13(8) | 0.45(-11) |
| Metropolis | -0.10(3) | 0.62(-12) | -0.06(8) | 0.52(11) | -0.21(-6) | -0.32(2) | -0.13(-5) | -0.39(4) |

| | | | | | | | |
|---|---|---|---|---|---|---|---|
| Hermanmo | 0.10(12) | 0.46(-2) | -0.12(-11) | 0.41(-1) | 0.11(12) | 0.18(-10) | -0.15(-12) | 0.3(0) |
| Vicksburg | -0.12(-8) | 0.63(1) | -0.09(-3) | -0.54(-6) | -0.20(-5) | -0.31(-9) | -0.15(-4) | -0.41(-7) |
| Alton | 0.16(-2) | 0.58(2) | 0.06(3) | 0.51(1) | -0.14(-4) | -0.27(-9) | -0.10(-12) | 0.40(12) |
| Obidos | -0.14(5) | -0.67(+4) | -0.08(9) | -0.56(+4) | -0.18(9) | -0.28(+8) | -0.13(9) | -0.39(+6) |
| Guaira | 0.15(8) | 0.64(+1) | 0.17(1) | 0.58(+2) | 0.17(-1) | 0.26(+4) | 0.17(1) | 0.41(+3) |
| Juazeiro | 0.14(-4) | 0.69(11) | 0.06(-11) | 0.59(10) | 0.16(-12) | 0.27(-4) | 0.10(-11) | 0.41(-3) |
| Calamar | -0.32(6) | -0.62(1) | -0.35(4) | -0.58(-1) | -0.31(1) | -0.34(-2) | -0.33(2) | -0.48(-1) |
| Albury | -0.12(-2) | -0.56(0) | -0.13(-2) | -0.50(-1) | -0.24(0) | -0.29(-4) | -0.16(-1) | -0.39(-2) |
| Swan Hill | 0.16(-10) | 0.59(+6) | -0.19(4) | -0.54(+1) | -0.30(0) | -0.35(-8) | -0.22(2) | -0.43(+2) |
| Taupo'Control | 0.09(-8) | -0.29(1) | 0.14(-12) | -0.29(0) | -0.09(0) | 0.16(-9) | 0.13(-12) | -0.23(0) |

Note :"()"indicates that the month of advance or delay, positive is the delay and the negative is advance.

**4 High and mid-latitudes oscillation index**

Atmospheric circulation is a supporter of all kinds of energy transformation in the earth atmosphere system, its anomaly directly affect the change of climate elements such as temperature, precipitation, wind, air pressure. NAO reflect an inverse transition relationship between two activity centers of the Azores high SLP and Icelandic low SLP change (Trenberth and Hurrell 1994). Thompson (2001) and Wallace (2000) pointed out that both AO and NAO are consistent nature and are two manifestations of the same thing in different sides. They actually reflect the strong and weak changes of the westerly wind belt in the middle latitude. The AAO have significant regulation to precipitation variability of many high latitude areas in the southern hemisphere (Gillett et al. 2006), and coupled general circulation model (CGCM) also can well simulate the AAO, further illustrate the AAO is an inherent feature of the climate system, is the basic regularities of the atmospheric circulation variation in the southern hemisphere (Schneider and Kinter 1994).

4.1NAOSSTD

The NAO is main modal of atmospheric circulation anomalies in the north Atlantic area, and is reflected ocean heat and power, sea and land distribution, which more show characteristics of regional atmospheric circulation, is the most important patterns affect the seasonal variation (Judah and Barlow 2005).NOAA provides NAO index defining with the standardized SLP difference between Iceland Reykjavik station and the Ponta Delgada station. The analysis results show that the correlation coefficient between NAO and global runoff is within 0.08-0.32(Tab.5).

The NAO is essentially thermal differences as a result of the Azores and Iceland, Iceland is a half permanent low-pressure center around 64 °N, the Azores high pressure center is formed near the Azores in the Atlantic, is also a high pressure of large subtropical permanent center near 35 °N, we defines the SSTD between Iceland and Azores sea area as the NAO SSTD. Associated with global runoff coefficient has a substantial increase, maximum correlation coefficient can reach 0.83, the highest of NAO is only 0.32, it prove that the impact of the NAO on the global hydrological climate is better represented by directly using the SSTD(Table 5).

Tab.5 The correlation coefficient between high and mid-latitudes oscillation index and global runoff

| Station | Max correlation coefficient(month) | | | | | | | |
|---|---|---|---|---|---|---|---|---|
| | NAO | NP | AO | AAO | NAOSSTD | NPOSSTD | AOSSTD | AAOSST |
| Hankou | -0.29(-6) | -0.67(+7) | -0.22(-6) | -0.13(-1) | 0.80(+1) | -0.61(-1) | 0.81(-1) | 0.80(-7) |
| Datong | 0.08(10) | 0.66(1) | 0.21(1) | -0.12(11) | 0.81(-1) | 0.60(5) | 0.87(-1) | 0.88(11) |
| maqu | 0.13(6) | -0.59(-5) | 0.11(2) | -0.08(-4) | 0.70(11) | -0.52(-1) | -0.74(-7) | -0.75(11) |
| Longmen | -0.32(-5) | -0.58(+8) | -0.23(-5) | -0.10(2) | 0.69(0) | -0.67(0) | 0.73(0) | 0.69(-6) |
| Haerbin | -0.27(+6) | -0.54(-4) | -0.26(+7) | 0.12(5) | 0.59(-1) | 0.50(0) | 0.66(12) | 0.65(6) |
| Wuzhou | 0.08(5) | -0.61(6) | -0.25(-6) | -0.13(10) | 0.74(10) | -0.65(10) | 0.81(11) | 0.79(-7) |

| | | | | | | | |
|---|---|---|---|---|---|---|---|
| Darband | 0.11(4) | -0.69(-6) | -0.32(-6) | 0.12(-6) | 0.81(-2) | -0.82(-1) | 0.90(-1) | 0.88(-7) |
| Davlaishwaram | 0.15(7) | -0.56(8) | -0.16(8) | -0.17(12) | 0.68(-1) | -0.72(0) | 0.78(0) | 0.75(-6) |
| Vijayawada | 0.15(7) | -0.56(7) | -0.14(8) | -0.18(12) | 0.68(-1) | -0.70(0) | 0.79(0) | 0.75(-6) |
| Belogore | 0.08(-7) | -0.61(-6) | -0.27(5) | -0.16(1) | 0.75(10) | -0.77(10) | 0.83(-2) | 0.80(-8) |
| P. Tunguska | -0.08(12) | -0.56(5) | -0.21(4) | -0.16(-11) | 0.70(-3) | -0.73(-3) | 0.77(-2) | 0.75(-8) |
| Kyusyur | 0.11(5) | -0.53(6) | -0.14(5) | -0.09(10) | 0.67(10) | -0.71(-2) | 0.71(-1) | 0.70(5) |
| Volgograd | 0.19(-8) | -0.51(5) | -0.10(4) | -0.17(9) | 0.61(9) | -0.79(9) | 0.72(-3) | 0.69(-9) |
| D-T. Severin | -0.15(1) | 0.49(1) | -0.27(2) | -0.12(11) | 0.51(8) | -0.47(8) | 0.56(8) | 0.55(-10) |
| Dongola | -0.31(-5) | -0.60(+8) | -0.23(-5) | -0.20(-11) | 0.75(0) | -0.81(0) | 0.83(-11) | 0.79(-5) |
| Malakal | -0.23(+9) | -0.56(+4) | 0.18(+3) | 0.16(-4) | -0.70(+7) | -0.51(+2) | 0.73(2) | 0.75(-4) |
| Khartoum | 0.24(-5) | 0.47(-10) | -0.08(9) | -0.20(-12) | 0.64(12) | -0.74(-12) | 0.70(-12) | 0.70(-6) |
| Bangui | -0.07(-12) | -0.71(9) | -0.27(8) | 0.13(8) | 0.83(-11) | -0.74(-11) | 0.91(-11) | 0.90(-5) |
| Ouesso | -0.31(-4) | -0.63(-3) | -0.28(+9) | -0.18(-1) | 0.70(+1) | -0.76(+2) | 0.79(-10) | 0.75(8) |
| Hope | 0.13(5) | -0.63(-6) | -0.17(-7) | 0.13(4) | 0.76(-2) | -0.78(-2) | 0.87(10) | 0.84(4) |
| The Dalles | 0.15(5) | -0.54(-7) | -0.11(-8) | -0.12(-2) | 0.69(-3) | -0.73(-3) | 0.75(-2) | 0.73(3) |
| Metropolis | 0.12(1) | -0.54(-10) | -0.15(-11) | 0.13(-1) | 0.63(-6) | -0.52(6) | 0.65(-6) | 0.66(12) |
| Hermanmo | 0.21(-3) | -0.33(-8) | 0.12(-2) | 0.09(1) | 0.44(-3) | -0.45(-3) | 0.53(-3) | 0.53(3) |
| Vicksburg | 0.11(4) | -0.53(-9) | -0.14(-9) | -0.12(-4) | 0.61(-5) | -0.53(7) | 0.67(-5) | 0.67(1) |
| Alton | 0.09(4) | -0.46(-8) | -0.14(3) | 0.14(1) | 0.56(-4) | -0.56(8) | 0.62(-4) | 0.62(2) |
| Obidos | 0.27(-2) | 0.54(-1) | 0.20(-2) | 0.17(3) | -0.69(-9) | 0.53(+2) | -0.78(-9) | -0.79(3) |
| Guaira | 0.23(+5) | -0.53(+1) | 0.20(+8) | 0.13(-11) | 0.62(-7) | -0.58(+6) | 0.72(-6) | 0.72(-12) |
| Juazeiro | 0.10(0) | -0.52(-11) | -0.22(1) | -0.15(6) | 0.63(5) | -0.53(6) | 0.71(-6) | 0.72(12) |
| Calamar | 0.14(9) | 0.50(-10) | 0.20(1) | 0.14(3) | -0.55(-6) | 0.60(6) | -0.64(7) | -0.60(-11) |
| Albury | -0.12(3) | 0.39(-10) | -0.09(8) | -0.10(8) | 0.53(-12) | -0.48(-11) | 0.57(1) | 0.57(7) |
| Swan Hill | -0.23(+7) | 0.44(+3) | -0.23(+8) | -0.22(1) | 0.54(0) | 0.46(+6) | -0.61(-11) | 0.62(-11) |
| Taupo'Control | 0.09(-5) | -0.31(1) | 0.07(-5) | -0.09(-4) | -0.30(7) | -0.24(-11) | -0.33(-4) | -0.32(-10) |

Note :"()"indicates that the month of advance or delay, positive is the delay and the negative is advance.

4.2 NPOSSTD

To describe the changes of the atmospheric activity center strength in the north Pacific region, Trenberth (1994) using the north Pacific region (30°N~65°N, 160°E~140°W) wide area weighted average of the SLP change defined NP. The correlation calculation between NP and global runoff is shown in Table 5, the correlation coefficients of NP and global runoff are higher than those of AO and NAO, and the maximum correlation coefficient is 0.69.

Wallace(1981) defined the NPO with the Aleutian low pressure area that is a semi-permanent activity center of deputy polar low-pressure in the northern Pacific near the Aleutian Islands, Hawaii high pressure zone is semi-permanent activity center in the Pacific Ocean (30°N) near the Hawaiian island, we define SSTD of the two key area as the NPOSSTD. The correlation calculation result with global runoff shows that most sites correlation coefficient than the original index (NP) significantly increase, the maximum is 0.81(Tab.5).

4.3 AOSST

The arctic oscillation (AO) is the most prominent modal of the SLP field change in the northern tropical area (north of 20°N), has obvious zonal symmetry on the space, its characteristic performance in terms of height field and SLP field, as a main activity center in polar and a corresponding around the polar, centered on 45°N, assumes the circular distribution, symbol with changes in the polar opposite activity center, also known as the circular mode (Wallace 2000; Clara Deser 2000). Zhao (2006) found that the AO exist an obvious core in the GIN Seas (Greenland Sea, Iceland Sea and Norwegian Sea)about the north of 70°N, the SLP change of the core can well represent the characteristics of the AO, we analyze correlation between SST of the north of 40 ˚N,50 ˚N,65 ˚N,70 ˚N and global seventy stations runoff, find the correlation coefficient of the 50 ˚N ring SST is the best, so select SST of north of 50 ˚N ring regional as the AOSST.NOAA provides the monthly mean AO, which is a standardized index of the average atmospheric pressure of the EOF first characteristic vector on the 1000hPa height field from the 20 ˚N to the North pole. The analysis of AO and global runoff correlation in table5, the correlation coefficient is 0.08-0.27.AnalyzingAOSSTrelevance with global runoff, the correlation coefficient also improve greatly, the maximum is 0.91, The AO affects not only the climate of the polar regions, but also the climate of the middle and low latitudes, which is an important atmospheric physical process with a global scale (Amir Givati 2013; Nagato and Tanaka 2012; Mao et al. 2011).

4.4 AAOSST

Thompson(2000) found AO and AAO have similarity and symmetry in the space, The Antarctic region also found a core region similar to that of the AO in the Antarctic region, the southern hemisphere high and mid-latitude SLP change between two atmospheric ring activities is a global scale seesaw structure. The standardization of the 40 ˚S and 65 ˚S zonal mean SLP difference as measure the AAO changes. Fan (2004) and Wang (2005) found that the AAO played a bridging role in the southern hemisphere's subtropical atmospheric circulation and the atmospheric convection in the western Pacific Oceanic continental region.

The AAO data provided by NOAA, which is a standardized index of the average atmospheric pressure of the EOF first characteristic vector on the 1000hPa height field from the 20 ˚S to the South pole. To verify with global runoff, the maximum correlation coefficient is only 0.22 (Tab. 5). With SST of 40 ˚S ring latitude to define the AAOSST, with the global runoff analysis the correlation find that the correlation coefficient is generally improved, the maximum can reach 0.90 (Tab.5), is almost strong function related, can be used as a climate factors using quantitative analysis of hydrology and climate.

**5 The equatorial Atlantic, Pacific and Indian Ocean joint oscillation index**

The SST of the western Pacific warm pool is global highest, which is the largest global atmospheric heat source, and the high temperature zone of warm pool is the corresponding atmospheric deep convection zone, which have triggered effect on ENSO events (Fedorov et al. 2015).Lau (1996) first put forward the mechanism of "atmosphere bridge", tropical Pacific SSTA changes through the tropical atmospheric bridge connected with other tropical oceans, ENSO mainly through zonal wind vertical shear rate, SST, Walker circulation and the western Pacific monsoon trough to influence the tropical cyclone activities(Gray 1984; Jones and Thorncroft 1998).There is a remarkable alternating seesaw effect between western Indian Ocean and western Pacific SST, the SST change with the IOD well known is very similar, is across the ocean dipole or oscillation, the western Indian-western Pacific ocean-atmosphere system and change have great influence on global weather and climate.

5.1 IPJOSSTD

According to the ENSO area of Pacific Ocean and the IOD area, Ju (2004) select two ocean relatively large SST zonal variable rate as two indicators area, that is western Indian Ocean and western Pacific SST Analyzed the difference, defines the WIPJO. Correlation analysis with global runoff, the correlation coefficient is higher than the standardization index, which reached 0.55 (table 6). Apart from Haerbin station, all other stations can be inspected by 95% confidence level, but do not have the meaning of quantitative analysis. This is because the Analyzed processing has the simplest filtering effect, which destroys the structure of the original data, resulting in a significant reduction in the correlation coefficient with global runoff.

Observation proved that the SST of equatorial WIPJO defined area has very obvious opposite variation characteristics, and on global climate have strong drive effect, due to improper analyzed difference index definition, cause the correlation coefficient are not significantly improved. On the basis of the definition of the SSTA difference by Ju (2004), the SSTD between the western Pacific and western Indian Ocean definition area is defined as the WIPJOSSTD. Correlation analysis with global runoff has been significantly improved, up to 0.83 (table 6).

Tab.6 The correlation coefficient between equatorial Atlantic, Pacific, Indian joint oscillation index and global runoff

| Station | Max correlation coefficient(month) | | | | | | Station | Max correlation coefficient(month) | | | | | |
|---|---|---|---|---|---|---|---|---|---|---|---|---|---|
| | WIPSSTA | WIPSSTD | APSSTA | APSSTD | IOD | NinoΔT | | WIPSSTA | WIPSSTD | APSSTA | APSSTD | IOD | NinoΔT |
| Hankou | -0.07(-4) | 0.78(+5) | -0.10(-5) | -0.82(+4) | 0.46(0) | -0.86(-7) | Khartoum | 0.41(-2) | 0.64(-11) | -0.11(-2) | -0.63(-1) | 0.64(1) | -0.72(-6) |
| Datong | 0.55(-2) | 0.77(-1) | -0.17(-5) | -0.83(-2) | 0.47(-1) | -0.80(-7) | Bangui | -0.53(-6) | -0.83(6) | -0.10(-2) | -0.83(-12) | 0.54(2) | -0.86(7) |
| maqu | -0.41(-8) | -0.71(-7) | 0.09(-12) | 0.69(-8) | -0.42(-7) | -0.71(5) | Ouesso | -0.1(-3) | 0.76(-6) | -0.09(7) | -0.67(0) | 0.57(3) | -0.70(-5) |
| Longmen | -0.09(+7) | 0.65(-8) | -0.12(-2) | -0.67(+1) | 0.52(-11) | -0.68(-6) | Hope | -0.54(-9) | 0.78(10) | -0.09(7) | 0.76(-9) | 0.66(11) | -0.82(-8) |
| Haerbin | 0.04(-10) | 0.65(-8) | -0.10(-5) | -0.64(-8) | -0.43(-7) | -0.63(5) | The Dalles | 0.50(7) | -0.66(-10) | -0.16(7) | -0.70(8) | 0.58(10) | -0.75(-9) |
| Wuzhou | -0.48(3) | 0.70(-2) | -0.12(-4) | -0.71(-3) | 0.51(-1) | -0.78(-8) | Metropolis | 0.41(5) | 0.59(6) | -0.11(-3) | -0.61(5) | 0.38(-5) | -0.66(0) |
| Darband | -0.55(-8) | -0.80(4) | -0.10(-3) | -0.78(-2) | 0.67(0) | -0.84(-8) | Hermanmo | -0.1(-10) | -0.51(-4) | 0.14(11) | 0.5(-2) | 0.35(-3) | -0.52(3) |
| Davlaishwaram | -0.43(-7) | 0.67(0) | -0.14(-2) | -0.67(-1) | 0.62(1) | -0.73(-6) | Vicksburg | 0.43(6) | 0.60(-5) | -0.09(-3) | 0.63(12) | -0.38(1) | -0.67(1) |
| Vijayawada | -0.45(-7) | 0.70(0) | -0.14(-2) | -0.69(-1) | 0.63(1) | -0.74(-7) | Alton | -0.38(1) | -0.56(1) | 0.13(-2) | 0.56(1) | 0.43(-3) | -0.61(2) |
| Belogore | -0.51(3) | 0.73(-2) | -0.18(-2) | 0.71(3) | 0.66(11) | -0.80(-8) | Obidos | -0.07(+7) | 0.67(+3) | 0.06(-12) | -0.72(-8) | -0.45(3) | 0.73(8) |
| P. Tunguska | -0.44(3) | 0.68(-2) | -0.23(8) | -0.65(-3) | 0.56(11) | -0.73(3) | Guaira | 0.15(+3) | 0.66(-2) | 0.15(7) | 0.69(+1) | 0.42(7) | -0.68(11) |
| Kyusyur | -0.42(3) | 0.63(-1) | -0.06(-4) | -0.59(-2) | 0.54(11) | -0.70(4) | Juazeiro | -0.45(-2) | 0.62(6) | 0.09(-4) | 0.65(11) | 0.42(6) | -0.69(11) |
| Volgograd | -0.47(2) | -0.62(-10) | 0.05(10) | 0.61(2) | 0.60(-2) | -0.72(3(3)) | Calamar | 0.53(0) | -0.59(-5) | -0.39(4) | -0.63(0) | -0.54(-5) | 0.58(1) |
| D-T. Severin | -0.50(9) | -0.56(1) | 0.08(12) | 0.49(1) | 0.35(-3) | -0.52(-10) | Albury | 0.42(-1) | 0.52(0) | 0.13(11) | -0.55(-1) | 0.39(1) | -0.57(-6) |
| Dongola | -0.1(-5) | -0.74(+1) | -0.13(-2) | -0.7(+1) | 0.64(1) | -0.80(-6) | Swan Hill | -0.19(+1) | -0.58(0) | -0.15(-1) | -0.6(0) | 0.44(1) | -0.62(-6) |
| Malakal | 0.15(-10) | 0.70(-6) | 0.15(-8) | -0.7(-1) | -0.44(-4) | -0.73(-5) | Taupo'Control | 0.29(9) | 0.32(1) | 0.10(8) | -0.31(1) | 0.26(1) | -0.29(-5) |

Note :"()"indicates that the month of advance or delay, positive is the delay and the negative is advance; Nino Δ T= Nino4- Nino 1+2.

5.2 APJOSSTD

The correlation of SSTD index than the SSTA index has greatly improved, so we direct use of the Atlantic and Pacific SSTD construct APJOSSTD. Based on the EOF method select two key SST area of the equatorial Atlantic and the equatorial Pacific that has a very significant seesaw phenomenon, the SST on the Pacific is generally higher than the Atlantic during the period of January to July, during August to December, it is opposite. The correlation coefficient between the APJOSSTD and global runoff is generally high, between 0.31 and 0.83. All stations are able to pass significant tests, and correlation coefficients can be used to qualitatively analyze climate change.

5.3 IOD and Nino4-Nino1+2

Saji (1999) based on two area SST reverse phase change on the Indian, using SSTA difference define the IOD, on the basis of these two area SSTD to redefine the IODSSTD, with global runoff correlation analysis show that the correlation coefficient is generally higher than that of standardized index and analyzed difference index, between 0.26 to 0.67. With Nino4 area and Nino1+2 area SSTD definition equatorial Pacific oscillation index, associated with global runoff analysis, in addition to Taupo 'Control station in Australia minimum of 0.29, other stations correlation coefficient are above 0.52, the highest reached 0.86, second only to AAOSSTD.

**6 Conclusions and discussion**

1. The ENSO phenomenon, which is a significant event in the interaction between the western Pacific warm and the eastern Pacific cold pool, in order to be able to better predict ENSO event, the researchers defined many index, which make data standardized processing, result in the coefficient between the index and global runoff is very low. Most of the correlation coefficient is lower than 0.1 between SOI and global runoff; the index of MEI and Nino monitoring area are slightly higher than SOI, which is mostly between 0.1-0.2, the maximum is 0.36, which cannot be used for quantitative analysis. AO, AAO and NAO have been further improved with global runoff correlation, most of which are around 0.2, with a maximum of 0.39, no substantial improvement; The NP is special, it is the average SLP in the northern Pacific Ocean, so the correlation is the highest among various indexes, and its correlation coefficient is comparable with the SSTD index.

2. The correlation coefficient between all indicators redefined with SST or SSTD and global runoff is greatly improved, and the ENSOSSTD up to 0.79, the correlation of Nino1+2 area is best, the maximum is 0.84, the Nino3 area up to 0.75, Nino4 area up to 0.47, Nino3.4 area up to 0.55; NAOSSTD is 0.83, and NPOSSTD is 0.82. AOSST is 0.81, AAOSST is 0.91.

3. The most correlation coefficient between AAOSST and global seventy stations runoff are above 0.5, the maximum can reach 0.91, which show that global climate change has reached high levels of response to climate factors, the four indexes defined with SSTD and Asian, African regional runoff response generally obvious, can be used for quantitative analysis of global climate and hydrology change.

Atmospheric oscillation and ocean-atmosphere theories can clear explain all kinds of hydrology and climate phenomenon with physical mechanism, and perfect the theory of climatology, greatly promote the development of the weather and climate prediction, and is widely used in agriculture, water resources, and fishery (Chen et al. 2007; Wheeler et al. 2000). However, its downside is also obvious that the correlation coefficient between various hydrological climate sequences is extremely low, reducing the credibility of the ocean-atmosphere coupling theory, the fundamental reason is that most of the index is standardized processing, the researchers deliberately nonlinear processing the time sequence of the indices, resulting in a significant reduction in the correlation coefficients, the nonlinear processing indexes only for climate diagnosis perhaps the problem is not big, if directly is used to predict calculation as a key parameter ,which will produce very big error. For the simplest analyzed calculation is also the most simple filtering processing, the correlation coefficient is greatly reduced. By redefining these indexes with SST or SSTD, such a problem would not be produced and could be used as a diagnostic indicator to calculate parameters.

Tab. 7 The correlation coefficient between the runoff of Hankou, Darband, Hope and Bangui station and each climate index

| Station | AAOSSTD | AOSSTD | NAOSSTD | Nino1+2 | Aleutian | Hawaii | Azores | Nino$\Delta T$ | WIPJOSSTD | APSSTD |
|---|---|---|---|---|---|---|---|---|---|---|
| Hankou | -0.83 | -0.762 | 0.705 | -0.689 | 0.82 | 0.849 | 0.836 | -0.689 | 0.667 | -0.515 |
| | 0.866 | 0.761 | 0.79(11) | 0.821 | -0.857 | -0.876 | -0.848 | 0.821 | -0.777 | 0.80(4) |
| | -0.88 | -0.765 | 0.804 | -0.836 | 0.867 | 0.882 | 0.861 | -0.836 | 0.775 | -0.82(2) |
| Darband | -0.67 | -0.731 | 0.474 | -0.488 | 0.679 | 0.701 | 0.718 | -0.488 | 0.629 | -0.282 |
| | -0.84 | -0.806 | 0.807(10) | 0.820(4) | 0.911 | 0.849 | 0.9 | 0.820(4) | -0.801 | 0.767(4) |
| | -0.84 | -0.813 | 0.807(2) | 0.823(8) | 0.916 | 0.851 | 0.897 | 0.823(8) | -0.79(8) | 0.777(8) |
| Hope | -0.45 | -0.469 | 0.228 | -0.276 | 0.364 | 0.466 | 0.421 | -0.276 | 0.317 | -0.043 |
| | -0.8 | -0.81 | 0.759 | -0.727 | 0.883 | 0.83 | 0.863 | 0.763(4) | 0.781 | 0.744 |
| | -0.8 | -0.802 | 0.758 | 0.785(8) | 0.883 | 0.822 | 0.862 | 0.785(8) | 0.765 | 0.757 |
| Bangui | -0.77 | -0.58 | 0.729 | -0.792 | 0.731 | 0.73 | 0.685 | -0.792 | 0.665 | -0.827 |
| | -0.91 | -0.773 | 0.808 | 0.841(7) | 0.92 | 0.893 | 0.88 | 0.841(7) | -0.83(6) | -0.724 |
| | 0.89(5) | -0.785 | 0.834 | 0.84(5) | 0.922 | -0.898(5) | 0.889 | 0.836(5) | -0.80(5) | -0.83(12) |

Note:( ) writing the month belongs to individual special advance and delay month.; Nino $\Delta T$= Nino4- Nino 1+2.

Before Correlation coefficient is very low, which is widely accepted by people, because all sorts of oscillation and coupled ocean-atmosphere theory can accurate explain various physical mechanism of the hydrology and climate phenomenon, construct theoretical system is strict self consistent. But with global runoff data correlation coefficient is extremely low, the various observations and the physical mechanism proved to be extremely significantly contact with atmospheric circulation, its low correlation coefficient cannot also support these theories, which is extremely abnormal obviously. This led us to reconstruct the various oscillation indexes by using the SSTD of the key area, and the correlation coefficient of global runoff data was greatly improved. Even more shocking the correlation between AAOSST, AOSSST, IPJOSSTD, APJOSSTD, Nino4-Nino1+2, Nino1+2 area, and even the Aleutian, Hawaii and Azores regional SST and the Hankou station of Yangtze river, Darband station of Ind, Hope station of Fraser, Bangui station of Oubangui generally exceeds 0.8 strong degree (Tab.7and Fig.1) , this particular phenomenon cannot be explained by the current atmospheric circulation theory. Because any physical internal elements of the earth system generally in the process of propagation attenuation, the geographic distance very far away the sea area SST and SSTD in the same region has such a high correlation coefficient of runoff, Proving that these relevant elements have a high resonance and resonance effect, there must be a common driving physical quantity, We do not think that the cause is within the earth system, but rather that the solar activity gives them the characteristic of a common cycle.

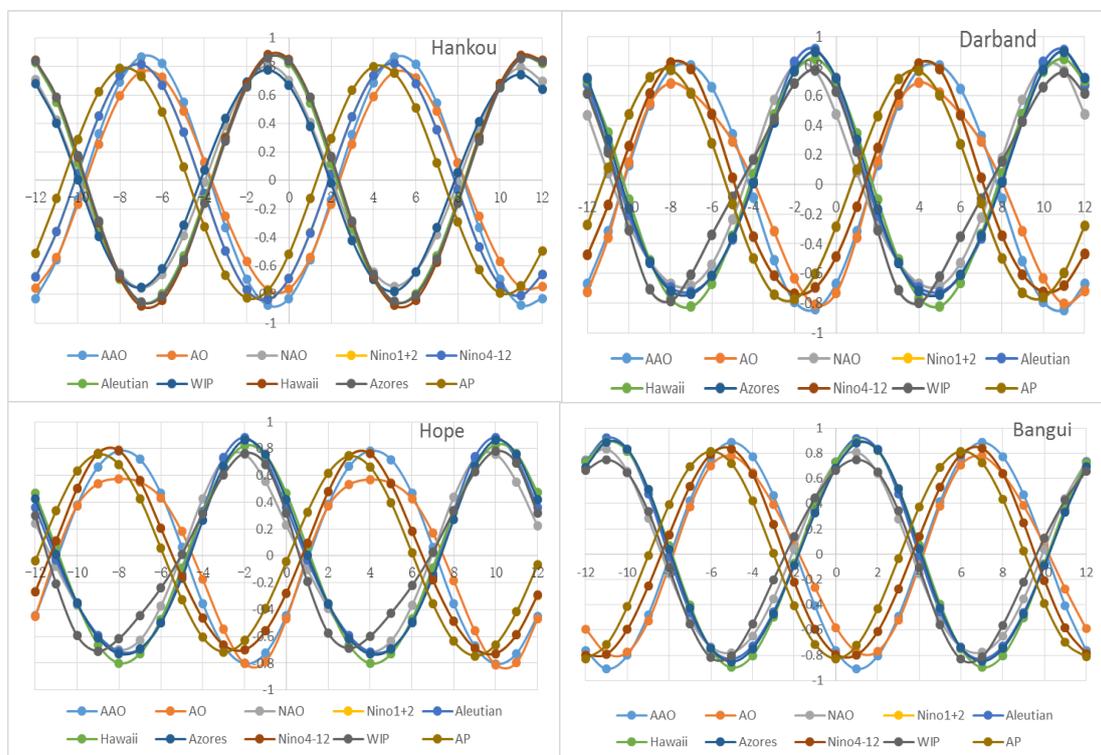

Figure 1: the correlation coefficients of the Hankou, Darband, Hope, Bangui runoff and the various indexes

As can be seen from figure1, the correlation of the above ten indexes is very good with the runoff of Hankou, Darband, Hope and Bangui station, and the circulation cycle of the six months is also very clear. AAOSST, AOSST, NINO1+2, Nino4-Nino1+2 and APJOSSTD are positive correlation, and the rest is negatively correlated. The first extremal value of late lag of Hankou and Darbandis in the fourth or fifth month, and the second is in the 10th or 11th month; The first extremal value of previous lag is in the first or second month, and the second is in the seventh or eighth month. The first extremal value of late lag of Hope station of Fraser is in the third or fourth month, and the second is in the ninth or 10th month; The first extremal value of previous lag is in the second or third month, and the second is in the eighth or ninth month. The first extremal value of late lag of Bangui station of Oubangui is in the zeroth or first month, and the second is in the sixth or seventh month; The first extremal value of previous lag is in the fifth or sixth month, and the second is in the 10th or 11th month. We can see figure 2, regardless of sun-earth distance and Hankou runoff or each climate index, have extremely strong circle circular motion characteristics, further proof that the cycle change of hydrological climate, are decided by sun-earth distance.

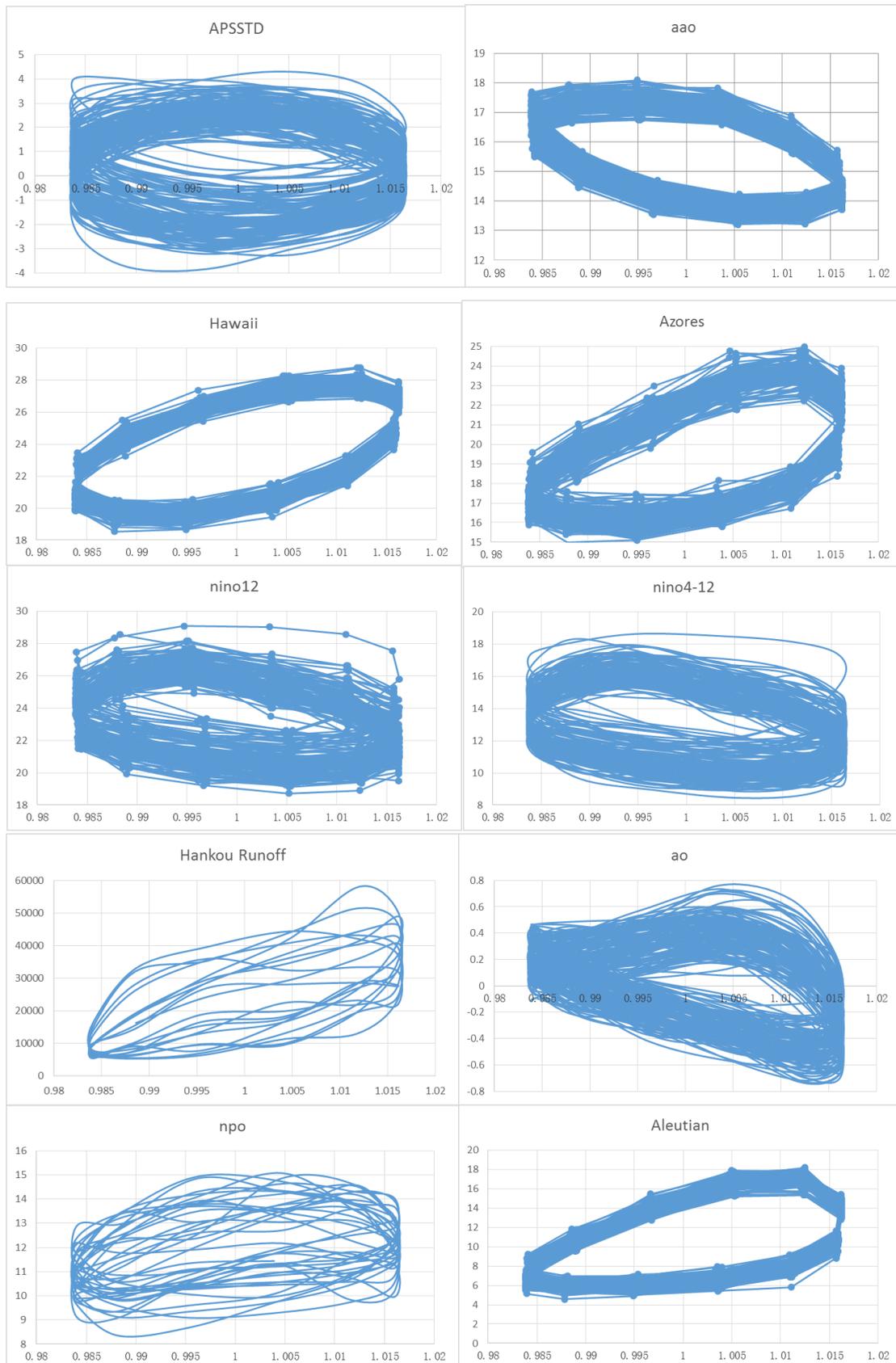

Fig. 2 Relationship of Hankou runoff and various climatic indicators

According to the calculation results, the SST and SSTD can more directly reflect thermal difference between the oceans and all kinds of coupled ocean-atmosphere and atmospheric oscillation phenomenon caused by the thermodynamic difference. We suggest use SST or SSTD directly instead of a standardized index and analyzed difference index, various standardized index is mainly used for climate diagnosis by meteorological department, and other interdisciplinary directly use these indicators to analyze correlation, using the low correlation coefficient to predict, let people think these physical relationship is very poor, the index can significantly improve the correlation coefficient by using the index of SST or SSTD directly, and then improve the prediction accuracy. At the same time the benefit of redefining the indicators of ocean-atmosphere coupling is: no change Nino phenomenon, coupled ocean-atmosphere, atmospheric circulation and weather shocks traditional classic theory, just confirmed that the astronomical factors driving the climate phenomenon change. The core of the problem is, the ocean as super filter of the earth system heat, with sun-earth distance changes have very strong resonance effect, we have found the astronomical parameters that can be use to accurate calculate the most regional SST and SSTD decades hundreds years, thoroughly solved various coupled ocean-atmosphere long-term prediction of long-term international problem, which is the root cause that we especially recommend SST or SSTD index.

**Acknowledgements**   We thank Zhangwen Liu, Lei Wang, Niming Zhang for helping on the manuscript. This work is major objectives of national research projects, was supported by the national natural science foundation.